\documentstyle[aaspp4,psfig]{article}

\begin{document}

\def\bc{\begin{center}}             \def\ec{\end{center}}
\def\be{\begin{equation}}           \def\ee{\end{equation}}
\def\lsim{\lower0.5ex\hbox{$\; \buildrel < \over \sim \;$}}
\def\gsim{\lower0.5ex\hbox{$\; \buildrel > \over \sim \;$}}
\def\rs{\mbox{$R_{\rm s}$}}         
\def\kc{\mbox{$k_{\rm c}$}}         
\def\dm{\mbox{$\dot{M}$}} 
\def\Bz{\mbox{$B_{\rm z}$}}
\def\bzs{\mbox{$B_{\rm z*}$}}
\def\Bf{\mbox{$B_{\phi}$}}
\def\bfs{\mbox{$B_{\phi*}$}}
\def\cs{\mbox{$c_{\rm s}$}}
\def\vf{\mbox{$v_{\phi}$}}
\def\ra{\mbox{$R_{\rm A}$}}
\def\rc{\mbox{$R_{\rm c}$}}
\def\rs{\mbox{$R_*$}}
\def\rin{\mbox{$R_{\rm in}$}}
\def\ro{\mbox{$R_{0}$}}
\def\os{\mbox{$\Omega_{\rm s}$}}
\def\om{\mbox{$\Omega$}}
\def\ok{\mbox{$\Omega_{\rm k}$}}
\def\oko{\mbox{$\Omega_{{\rm k}0}$}}
\def\ws{\mbox{$\omega_{\rm s}$}}
\def\wsc{\mbox{$\omega_{\rm c}$}}
\def\no{\mbox{$N_{0}$}}
\def\nin{\mbox{$N_{\rm in}$}}
\def\nout{\mbox{$N_{\rm out}$}}
\def\gm{\mbox{$\gamma$}}
\def\gms{\mbox{$\gamma_*$}}
\def\gs{\mbox{${\rm g}\,{\rm s}^{-1}$}}
\def\dmdr{\mbox{$\frac{{\rm d}\dm}{{\rm d}R}$}}
\def\dodr{\mbox{$\frac{{\rm d}\om}{{\rm d}R}$}}

\title{Disk Accretion onto Magnetized Neutron Stars: The Inner Disk Radius 
      and Fastness Parameter}

\author{
X.-D. Li and Z.-R. Wang
\\[12pt]  
%
Department of Astronomy, Nanjing University, Nanjing 210093, P. R. China
}

\begin{abstract}
It is well known that the accretion disk around a magnetized compact star
can penetrate inside the magnetospheric boundary, so the magnetospheric
radius $\ro$ does not represent the true inner edge $\rin$ of the disk;
but controversies exist in the literature concerning the relation between 
$\ro$ and $\rin$. In the model of Ghosh \& Lamb, the width of the boundary 
layer is given by $\delta=\ro-\rin\ll\ro$, or $\rin\simeq\ro$,  while 
Li \& Wickramasinghe recently argued that $\rin$ could be significantly 
smaller than $\ro$ in the case of a slow rotator. Here we show 
that if the star is able to absorb the angular momentum of disk plasma
at $\ro$, appropriate for binary X-ray pulsars,
the inner disk radius can be constrained by $0.8\lsim \rin/\ro\lsim 1$, and 
the star reaches spin equilibrium with a relatively large value 
of the fastness parameter ($\sim 0.7-0.95$).  For accreting neutron stars in
low-mass X-ray binaries (LMXBs), $\ro$ is generally close to the stellar
radius $\rs$ so that the toroidal field cannot transfer the spin-up torque
efficiently to the star. In this case the critical fastness parameter 
becomes smaller, but $\rin$ is still near $\ro$.

\end{abstract}


\section{Introduction}

Interaction between a magnetized, rotating star and a surrounding 
accretion disk has been of continued interest since the discovery of X-ray 
pulsars in binary systems (Giacconi et al. 1971, etc). Its solution may 
hold the key to understanding fundamental problems such as structure and 
dynamics of magnetized accretion disks, angular momentum exchange between 
the disk and the star, etc, in a variety of astrophysical systems, including 
cataclysmic variables, X-ray binaries, T Tauri stars and active galactic 
nuclei. A detail model for the disk-star interaction was first developed by 
Ghosh \& Lamb (1979a, 1979b), who argued that the stellar magnetic field 
penetrates the disk via Kelvin-Helmholtz instability, turbulent diffusion 
and reconnection, producing a broad transition zone joining the unperturbed 
disk flow far from the star to the magnetospheric flow near the star. The 
transition zone is composed of two qualitatively different regions, a broad 
outer zone where the angular velocity is nearly Keplerian and a narrow 
inner zone or boundary layer where it departs significantly from the 
Keplerian value.  The total torque $N$ exerted on the star with mass $M$
is then divided into two separate components:
\be N=\nin+\nout, \ee
where the subscript represents the appropriate zone in the disk. The 
``material" torque of the (inner) accretion flow is
\be \nin\simeq \no=\dm\ro^2\oko, \ee
where $\oko\equiv (GM/\ro^2)$ is the angular velocity of disk plasma at the 
magnetospheric radius $\ro$ (G is the gravitation constant), and $\dm$ is 
the accretion rate. Here we have adopted cylindrical coordinate $(R,\phi,z)$ 
centered on the star and the disk is assumed to be located on the $z=0$ 
plane, which is perpendicular to the star's spin and magnetic axes. The 
torque in the outer zone supplied by the Maxwell stress is given by
\be \nout=-\int_{\ro}^{\infty}\Bf\Bz R^{2}\,{\rm d}\,R, \ee
where $\Bf$ is the azimuthal field component generated by shear motion 
between the disk and the vertical field component $\Bz$.
The torque $\nout$ can be positive and negative depending on the value of 
the ``fastness parameter" $\ws\equiv \os/\oko=(\ro/\rc)^{3/2}$,
where $\os$ is the angular velocity of the star and 
$\rc\equiv (GM/\os^2)^{1/3}$ is the corotation radius.

Inside the magnetospheric boundary, the magnetic stress dominates the 
rotation of disk plasma, and forces it to corotate with the magnetosphere
at the radius $\rin$\footnote{According to Spruit \& Taam (1990), the 
accreting gas, nearly corotating with the star, could drift further inward 
across the field by an interchange instability.}.  To understand how far 
the disk can extend into the stellar magnetosphere and
how the torque is transferred to the star, it is necessary to obtain the 
solutions of the structure of the magnetosphere.  Ghosh \& Lamb (1979a) 
suggested that $\ro$ and $\delta$ be constrained by the following relation
\be (R\Bz\Bf/4\pi)_02\pi\ro2\delta\simeq \dm(R\Delta\vf)_0 \ee
(the subscript 0 denotes quantities evaluated at $R=\ro$),
which follows the equation of angular momentum conservation,
though contribution from gas funnel flow to angular momentum transfer has 
not been included. Ghosh \& Lamb (1979a) also assumed a priori
that $\delta$ corresponds to the electromagnetic screening length, and is 
much smaller than $\ro$. To testify the self-consistency of this assumption, 
from equation (4) 
we derive $\delta\simeq (\ro/\ra)^{7/2}\ro$ when $\ws\ll 1$, where 
$\ra$ is the conventional Alf\'ven radius in spherical accretion. 
If $\ro\simeq 0.5\ra$ as obtained by Ghosh \& Lamb (1979a), then indeed
$\delta\ll\ro$. However, if $\ro\simeq\ra$ as suggested by some recent 
investigations (e.g., Arons 1993; Ostriker \& Shu 1995; Wang 1996; Li 1997),
$\delta$ will be as large as $\ro$. 

Li, Wickramasinghe \& R\"{u}diger (1996) have presented a class of funnel 
flow solutions in which the angular momentum is carried by the matter. 
These authors argue that for an ideally conducting star, the stellar 
boundary condition requires the toroidal field be very small and most of the 
angular momentum of the matter in the funnel be propagated back to the 
disk rather to the star. Li \& Wickramasinghe (1997) show that
a consequence of this result is that the true inner disk radius $\rin$
could be much smaller than $\ro$, with $\rin/\ro$ as low as $\sim 0.1-0.2$ 
in slow rotator case, and the values of the critical
fastness parameter at which the net torque vanishes would also change. 
This ``torqueless" suggestion was criticized by Wang (1997), who points out 
that a small toroidal field is enough to produce a significant
torque on the star, and that the viscous stress is unable to remove 
angular momentum back outwards from the inner edge of the accretion disk.

This paper is organized as follows. In section 2, based on Wang's (1997)
arguments, we calculate the lower limit of $\rin$ that is required for
efficient angular momentum transfer, and find it to be at least 
$\sim 0.8\ro$,
implying that $\ro$ is a good indicator of the inner edge of the disk.
In section 3 we discuss the possibility of inefficient angular momentum
transfer, which is likely to occur in LMXBs, and the implications for the
resulting variation in the fastness parameter.  We conclude in section 4.

\section{The inner disk radius}

An estimation of $\ro$ can be obtained by setting the magnetic stress equal
to the rate at which  angular momentum is removed in the disk (Wang 1987)
\be (R^{2}\Bf\Bz)_0=[\dm\frac{{\rm d}(R^2\om)}{{\rm d}R}]_0, \ee
where $\om=\om(R)$ is the angular velocity of disk plasma.
Following Li \& Wang (1996), we assume that $\om(R)$ reaches its maximum at 
$R=\ro$, i.e, 
$({\rm d}\om/{\rm d}R)_0=0$, with $\om(\ro)\simeq\oko$, and rewrite equation
(5) as
\be \frac{\Bf_0\Bz_0}{\dm(GM\ro)^{1/2}}=-\frac{2}{\ro^3}. \ee
The torque $\nin$ in equation (1) in fact contains two components: 
the real material torque $N_{\rm f}$ in the funnel flow, and the
magnetospheric torque $N_{\rm mag}$ arising from the shearing motion between 
the corotating magnetosphere and the non-Keplerian disk boundary layer, i.e.,
\begin{eqnarray}
    \nin &=& N_{\rm f}+N_{\rm mag} \nonumber \\
         &=& \int_{k\ro}^{\ro}\dmdr R^2\om\,{\rm d}\,R
             -\int_{k\ro}^{\ro}\Bf\Bz R^{2}\,{\rm d}\,R, 
\end{eqnarray}
where $k=\rin/\ro$. 

We first consider the second term on the righ-hand side of
equation (7). In estimating the toroidal magnetic filed $\Bf$, we
assume as usual that $\Bf$ is generated by the shearing between the disk and
the stellar magnetosphere, and the growth of $\Bf$ is limited by diffusive
decay due to turbulent mixing within the disk (e.g., Campbell 1992; Yi 1995;
Wang 1995)
\be \frac{\Bf}{\Bz}\propto \frac{\os-\om}{\om}. \ee
Several different phenomenological descriptions of disk-magnetosphere
interaction (Wang 1995) cause little practical differences to our conclusions.
Assuming that $\Bz$ follows a dipolar field and using equations (6) and (8), 
we may write the magnetospheric spin-up torque into dimensionless form as
\be \xi_{\rm mag}=\frac{N_{\rm mag}}{\no}
                 =\frac{4\ws^2}{3(1-\ws)}\int_{k^{3/2}\ws}^{\ws}
                 [1-(\os/\om)]y^{-3}\,{\rm d}\,y, \ee
where $y=(R/\rc)^{3/2}$. 

In the first term  on the righ-hand side of equation (7), 
$\dmdr$
denotes the vertical mass loss rate from the boundary layer. This can be 
described by the equation (e.g., Ghosh \& Lamb 1979a)
\be \dmdr=4\pi R\rho\cs g(R), \ee
scaling the vertical flow velocity from the boundary layer in terms of
the local sound speed $\cs$ and introducing a ``gate" function $g(R)$
describing the radial profile of mass loss 
out of the boundary layer ($\rho$ is the mass 
density). To keep things simple, we assume $g(R)=0$ outside $\ro$,
and $g(R)=1$ when $\rin\le R\le\ro$. Assuming that the disk plasma around
$\ro$ is thermally supported and optically thick to Thomson scattering
(appropriate for a binary X-ray pulsar), from the $\alpha$-disk model of
Shakura \& Sunyaev (1973), we obtain
$\dmdr\propto R^{-1}$ or $\dmdr=\frac{\dm}{(-{\rm log}k) R}$, which leads to 
\be \xi_{\rm f}=\frac{N_{\rm f}}{\no}= \frac{1}{(-{\rm log}k)}
    \int_{k\ro}^{\ro}(\frac{\om}{\oko})\frac{R}{\ro^2}{\rm d}R. \ee
For generality, we have also assumed $\dmdr=const$,
and found that the results do not change significantly.

The value of $k$ can be obtained by setting 
$\xi=\xi_{\rm f}+\xi_{\rm mag}=1$.
From equations (9) and (11) it can be seen that the magnitude of $k$ depends 
on the detail profile of $\om(R)$ between $\rin$ and $\ro$, which is unknown
before solving the magnetohydrodynamics (MHD) equations.  A lower limit
of $k$, however, can be estimated in the following way. Because of the steep
radial dependence of the stellar magnetic field ($\Bz\propto R^{-3}$), 
the slope of the angular velocity $\dodr$ should increase from 0 at $\ro$ to 
a large value at $\rin$. This requires that, in the $\om(R) {\rm vs.} R$ 
diagram (Figure 1a) $\om(R)$ should always lie beyond the dashed line which 
represents a linear relation between $\om$ and $R$. As we know from equations 
(9) and (11), the larger $\om$, the larger the $\xi$, and the smaller 
the $\delta$ 
required. So the smallest $k$ corresponds to the extreme case that $\om(R)$
decreases linearly from $\oko$ to $\os$ between $\ro$ and $\rin$, i.e.,
\be \om(R)=aR+b, \ee
where
\be a=\frac{\oko}{\ro}\frac{(1-\ws)}{(1-k)}, \ee
and
\be b=\oko\frac{(\ws-k)}{(1-k)}. \ee

In Figure 1b we plot the calculated values of $k$ as a function of the 
fastness parameter $\ws$ in the dashed curve. It is seen that $k$ naturally
reaches 1 when $\ws=1$, but remains $\sim 0.8$ within a large range of 
$\ws$.  The solid curves in Figure 1a and 1b represent an example with more 
practical boundary conditions, which are always located above the dashed 
curves.

\begin{figure}[htbp]
  \centerline{\psfig{figure=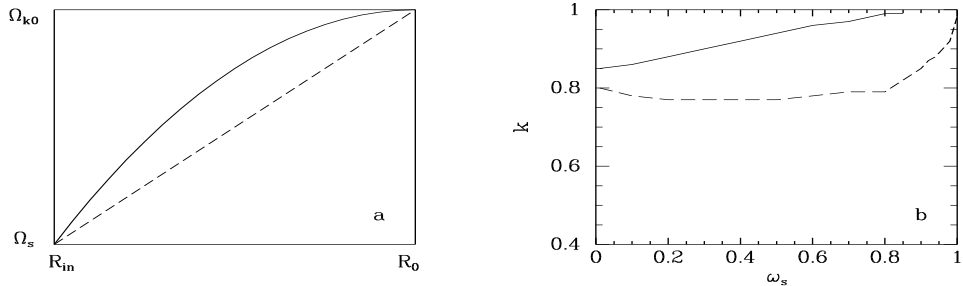,height=7.8cm,width=16.0cm,angle=-90}}
  \caption{(a) The radial profile of the angular velocity $\om(r)$ of disk
  plasma in the boundary layer. The dashed line denotes a linear decrease of
  $\om(R)$ with $R$, $\om-R$ relations above which are physically allowed.
  The solid curve represents one of the practical cases. (b) The radius ratio
  of $k=\rin/\ro$ as a function of the fastness parameter $\ws$. The solid and
  dashed curves correspond to those in (a), respectively. The latter puts a
  lower limit on $k$.}
\end{figure}

\section{The fastness parameter}

Our calculations in last section are based on the assumption that a torque
of order $\dm\ro^2\oko$ can be efficiently transmited to the star by both 
the gas funnel flow and the magnetic stress. It has been suggested that 
(Shu et al. 1994; Ostriker \& Shu 1995; Li, Wickramasinghe, \& R\"udiger 
1996) a star may accrete matter from a magnetically truncated Keplerian 
disk without experiencing any spin-up torque, because it would require the 
surface magnetic field to have a large azimuthal twist which would lead to 
dynamical instabilities. Wang (1997) estimated the azimuthal pitch $\gms$ at 
the stellar surface required to transmit this torque to be
\be
\gms\equiv\frac{\bfs}{\bzs}\simeq \frac{\xi}{2^{3/2}}
          (\frac{\rs}{\ro})^{3/2}\frac{\ro}{\delta},
\ee
where the subscript asterisk denotes quantities at the stellar surface. 
Since MHD instabilities are expected to occur only when $|\gms|\gsim 1$,
one sees that, with $\delta\lsim 0.2\ro$ and $\xi\simeq 1$, $\gms$ could be 
much less than unity if $\rs\ll \ro$. This condition is satisfied by 
binary X-ray pulsars, which generally
possess surface magnetic fields as strong as $10^{12}-10^{13}$ G, and hence
$\ro\sim 10^8-10^9$ cm with the mass accretion rates of $10^{16}-10^{18} \gs$,
much larger than $\rs\sim 10^6$ cm. Thus the standard accretion disk models
based on Ghosh \& Lamb (1979a, 1979b) are suitable for these systems, and
the value of the critical fastness parameter $\wsc$, according to 
Wang (1995) and Li \& Wang (1996), may lie between $\sim 0.7$ and 0.95.

There could exist possibility of an accreting star unable to
absorb the excess angular momentum at $\ro$. This may occur in accreting
neutron stars in LMXBs, in which the condition $|\gms|\ll 1$ may not hold
unless $\xi\sim 0$, because the surface field strengths in these
neutron stars are so low ($10^8-10^9$ G or lower) that the accretion disk 
usually extends close to the star's surface, i.e., $\ro\sim \rs$.
Since MHD instabilies prevent transmission of the torque of $\nin$,
in this case the star only experiences the magnetic torque $\nout$, and
the critical fastness parameter decreases to $\sim 0.4-0.6$ 
(Li \& Wickramasinghe 1997).  Note that the angular momentum of the 
material at $\ro$ cannot be transported from the boundary layer to the outer 
parts of the disk by either
viscous stress or a stellar magnetic field threading the inner region of the 
disk (Wang 1997), it is more likely to be removed by a magnetocentrifugal
wind. During wind mass loss, some angular momentum of the star could also be
lost, equivalent to an extra braking torque exerted on the star,
so that the critical fastness parameter would be even smaller.

Observational evidence for the above arguments comes from
kiloHertz quasi-periodic oscillations (kHz QPOs) 
recently discovered with the {\em Rossi X-ray Timing Explorer} (RXTE) 
in some 16 neutron star LMXBs (see van der Klis 1997 for a review).
These kHz QPOs are characterized by their
high levels of coherence (with quality factors $Q\equiv \nu/\Delta \nu$
up to $\sim 200$), large rms amplitudes (up to several $10\%$), and wide
span of frequencies ($\sim 300-1200$ Hz), which, in most cases, are strongly
correlated with X-ray fluxes. In many cases, two simultaneous kHz peaks are
observed in the power spectra of the X-ray count rate variations, with the
separation frequency roughly constant (e.g., Strohmayer et al. 1996;
Ford et al. 1997; Wijnands et al. 1997). Sometimes a third kHz peak is
detected in a few atoll sources during type I X-ray bursts at a frequency
equal to the separation frequency of the two peaks (Strohmayer et al. 1996;
Ford et al.
1997) or twice that (Wijnands \& van der Klis 1997; Smith, Morgan \& Bradt
1997). This strongly suggests a beat-frequency interpretation, with the
third peak at the neutron star spin frequency (or twice that), the upper
kHz peak at the Keplerian orbital frequency at a preferred radius around the
neutron star, and the lower kHz peak at the difference frequency between
them\footnote{The applicability of such a model is less clear when the
difference frequency is not constant as in the Z source Sco X-1 and possibly 
the atoll source 4U 1608-52; see van der Klis et al. (1996) and 
M\'endez et al. (1998).}.
If the upper kHz QPOs are associated with the orbital frequency at the 
magnetospheric radius or the sonic point in the accretion disk 
(Strohmayer et al. 1996; Miller et al. 1998; see also Lai 1998), one would
expect that the disk can extend towards the marginally stable orbit of 
a neutron star. 

As pointed by Zhang et al. (1997) and White \& Zhang (1997), these neutron 
stars may have accreted a considerable amount of mass, so they possibly
have reached the equilibrium spin. The measured spin and QPO frequencies 
of these neutron stars reveal the critical fastness parameter $\wsc$ lying 
between $\sim$ 0.2 and 0.7.  Additionally,
considering the spin and orbital evolution in low-mass binary pulsars which
are thought to originate from LMXBs, Burderi et al. (1996) also found a
small value of $\wsc\sim 0.1$ from the observed relation between spin period, 
magnetic field and orbital period. Though subject to both observational and 
theoretical uncertainties, these results do indicate a smaller fastness 
parameter than predicted by current theory for binary X-ray pulsars.

\section{Conclusions}

We summarize our results as follows: (1) The inner disk radius $\rin$ is 
always close to the magnetospheric radius $\ro$. (2) For binary X-ray 
pulsars, the critical fastness parameter lies in the rang $0.71-0.95$; 
neutron stars in LMXBs may not absorb angular momentum flux at $\ro$ as 
efficiently as X-ray pulsars, resulting in a critical fastness parameter 
considerably smaller than in X-ray pulsars.

\acknowledgements
This work was supported by National Natural Science Foundation of China.

 

\end{document}